\documentclass[11pt]{article} 
\usepackage{amsfonts,amscd,amsmath,amssymb,graphicx,color}

\def\ep{\text{e}}

\def\zc{z_{\text{\tiny c}}}
\def\g{\mathfrak{g}}
\def\s{\mathfrak{s}}
\def\zt{z_{\text{\tiny T}}}
\def\T{T_c}

\title{Renormalized Field Strength Correlators in $SU(N)$ Gauge Theory and Gauge/String Duality}
\author{Oleg Andreev\thanks{Also at Landau Institute for Theoretical Physics, Moscow.}
\\ \\
{\it Arnold Sommerfeld Center for Theoretical Physics, LMU-M\"unchen,} \\
{\it Theresienstrasse 37, 80333 M\"unchen, Germany}}
\date{}

\begin{document} 

\vspace{-8cm} 
\maketitle 
\begin{abstract} 
We use gauge/string duality to {\it analytically} evaluate correlation lengths of the renormalized field strength correlators in pure Yang-Mills theories
at zero and finite temperature.
 \end{abstract}

\vspace{-10cm}
\begin{flushright}
LMU-ASC 31/10
\end{flushright}
\vspace{9cm}


\section{Introduction}
Understanding the infrared behavior of gauge theories from first principles is a longstanding problem that offers perhaps the best hope of eventually understanding all the mysteries of quantum chromodynamics (QCD). It is well known that an important class of gauge invariant correlation functions in the QCD vacuum is constructed from field strengths and Wilson lines. These correlators do play an important role in several areas of QCD 
including models of stochastic confinement of color, high energy scattering, heavy quarkonium systems, etc.\footnote{The literature on the field strength correlators is vast. For a review, see, e.g., \cite{review} and references therein.}

The simplest field strength correlator in four-dimensional Euclidean space is defined by

\begin{equation}\label{F2}
\text{D}_{\mu\nu ,\rho\tau}(x)=\langle\,\text{tr}\bigl[
	G_{\mu\nu}(0)U_P(0,x)G_{\rho\tau}(x)U_P(x,0)
	\bigr]\,\rangle
	\,.
\end{equation}
Here $x^\mu$ are the Euclidean coordinates, the trace is over the fundamental representation, $G_{\mu\nu}$ is a field strength of the gauge field $A_\mu$ and $U_P(x,0)$ is a path-ordered Wilson line. The last is defined as $U_P(x_1,x_2)=P\exp\bigl[ig\int_0^1ds \frac{dx^\mu}{ds}\,A_\mu(x(s)) \bigr]$, where $s$ is a parameter of the path running from $0$ at $x=x_1$ to $1$ at $x=x_2$ and $g$ is a gauge coupling constant.\footnote{In what follows we omit the indices when it is clear from the context.} The paths are defined as straight lines.

If one considers not a field strength but a field with different color quantum numbers, then the construction of correlators is changed. For instance, a field transforming in the fundamental representation of the gauge group will produce a two-point correlator 

\begin{equation}\label{f2}
\Psi(x)=\langle\,\bar q (0)U_P(0,x) q(x)\,\rangle
	\,,
\end{equation}
with $q$ the field in the fundamental representation of $SU(N)$ and $\bar q$ its conjugate. In general, there are many 
possible generalizations of the correlators by choosing different (curved) contours for the Wilson lines, or by inserting operators along the contours.

It is believed that for large separation of the field strength operators the two-point correlator falls off exponentially \cite{review}

\begin{equation}\label{F2large}
\text{D}_{\mu\nu ,\rho\tau}(x)\sim {\text P}_{\mu\nu,\rho\tau}\,\ep^{-\frac{r}{\lambda}}
	\,,
\end{equation}
where ${\text P}_{\mu\nu,\rho\tau}$ is a kinematical factor, $r=\sqrt{(x^\mu)^2}$, and $\lambda$ is a correlation length. This is the case also for 
$\Psi$, in fact

\begin{equation}\label{f2large}
\Psi(x)\sim\ep^{-\frac{r}{\xi}}
	\,,
\end{equation}
with a correlation length $\xi$.

Until recently, the lattice formulation, still struggling with limitations and system errors, was the main computational tools to deal with non-weakly coupled gauge theories. The field strength correlators were also intensively studied (for a brief review, see \cite{giacomo-rev}). The situation changed drastically with the invention of the AdS/CFT correspondence \cite{malda} that resumed interest in another tool, string theory. The original duality was for conformal theories, but various perturbations (deformations) produce gauge/string duals with a mass gap, confinement, chiral symmetry breaking, etc \cite{rev-ads}.

In this note we continue a series of recent studies devoted to a search for an effective string description of pure gauge theories. In \cite{az1}, the 
model was presented for computing the heavy quark potential at zero temperature. Subsequent comparison \cite{white} with the available lattice data has made it clear that the model should be taken seriously. A non-trivial cross check for this model \cite{az4}, which checked the phenomenological value of the gluon condensate \cite{svz}, was also carried out. Later, the model was extended to finite temperature. The results obtained for the spatial string tension \cite{az2} and the expectation value of the renormalized Polyakov loop \cite{p-loop} in the deconfined phase are remarkably consistent with the lattice, too. As is known, QCD is a very rich theory supposed to describe the whole spectrum of strong interaction phenomena. The question naturally arises: How 
well does such an effective string description capture other aspects of quenched QCD? Here, we address the issue of computing the field strength correlators in an {\it analytical} way as an important step toward answering this question.\footnote{To our knowledge, there have been no studies (numerical or analytical) of this issue from the viewpoint of AdS/CFT, or gauge/string duality, in the literature.}

Before proceeding to the detailed analysis, let us set the basic framework. As in \cite{az1,az4}, we take the following ansatz for the five-dimensional background geometry

\begin{equation}\label{metric}
ds^2={\cal G}_{NM}dX^N dX^M=
R^2 w
\left(dt^2+d\vec{x}^2+dz^2\right)
\,,\qquad
w(z)=\frac{\ep^{\s z^2}}{z^2}
\,
\end{equation}
to describe a pure gauge theory at zero temperature. The metric \eqref{metric} is that of a deformed $\text{AdS}_5$ space, where $\s$ is a deformation parameter whose value can be fixed, for example, from the heavy quark potential. We take a constant dilaton and discard other background fields.

When we go to the deconfined phase, we consider the five-dimensional geometry of \cite{az2,p-loop}

\begin{equation}\label{metricT}
ds^2=\mathfrak{G}_{NM}dX^NdX^M=
R^2 w
\left(f dt^2+d\vec{x}^2+\frac{1}{f}dz^2\right)
\,,\qquad
f(z)=1-\bigl(\tfrac{z}{\zt}\bigr)^4
\,,\phantom{=\frac{\ep^{\s z^2}}{z^2}}
\end{equation}
which represents a deformed Schwarzschild black hole in $\text{AdS}_5$ space. Here $\zt$ is related to the Hawking temperature $T =1/(\pi\zt)$ whose dual 
description is nothing but the temperature of gauge theory. 


\section{Calculating the Correlators at Zero Temperature}

To begin with, we need to find a recipe for computing the field strength correlators within the gauge/string duality. The strategy for finding it is as 
follows. In discussing the Wilson loops \cite{malda-rey}, one first chooses a contor ${\cal C}$ on a four-manifold which is the boundary of a five-dimensional manifold. Next, one has to study fundamental strings on this manifold such that the string world-sheet has ${\cal C}$ as its boundary. 
The expectation value of the Wilson loop is schematically given by the world-sheet path integral $\langle\, W({\cal C})\,\rangle=\int DX\,\ep^{-S_w}$, 
where $X$ denotes a set of world-sheet fields and $S_w$ is a world-sheet action. In principle, the integral can be evaluated approximately in
terms of minimal surfaces that obey the boundary conditions. The result is written as $\langle\,W({\cal C})\,\rangle=\sum_n w_n\ep^{-S_n}$, 
where $S_n$ means a renormalized minimal area (in string units) whose weight is $w_n$. 
\begin{figure}[ht]
\centering
\includegraphics[width=5.8cm]{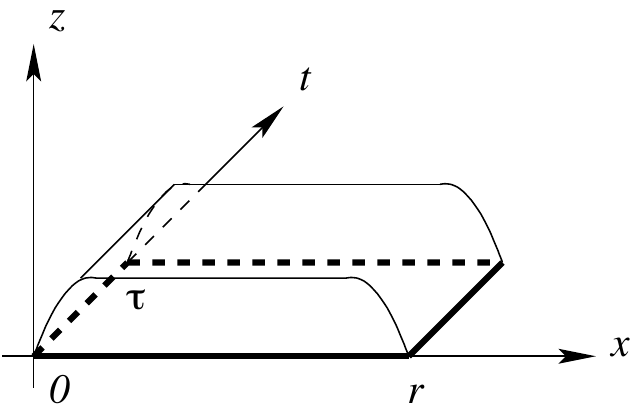}
\hfill
\includegraphics[width=5.8cm]{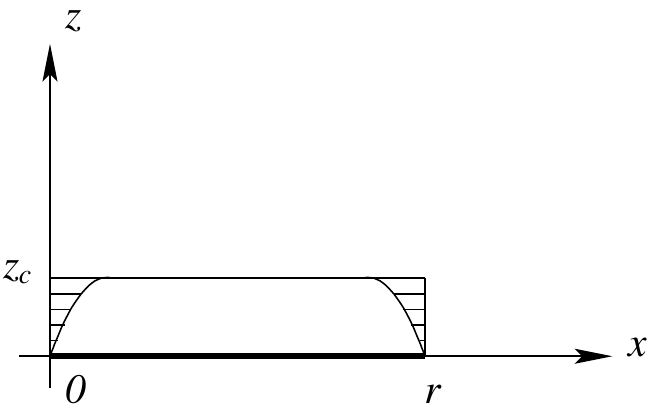}
\caption{\small{Left: A minimal surface for a rectangular loop ${\cal C}$ indicated by thick lines along the $(t,x)$-axes. It includes 
a lateral surface whose area is proportional to $\cal T$ and two identical end-surfaces with areas independent of ${\cal T}$. 
Right: The end-surface in the $x$-$z$ plane. It is bounded by a curved profile of a static string stretched between quark sources set at $x=0,r$ and a straight Wilson line indicated by a thick line along the $x$-axis. Note that in the model we are considering a gravitational force prevents the string from getting deeper than $\zc=1/\sqrt{\s}$ into $z$ direction \cite{az1}.}} 
\end{figure}

In the case of a rectangular Wilson loop living on the boundary ($z=0$) of five-dimensional space \eqref{metric} a minimal surface corresponding to a static string cofiguration is sewn together from three surfaces, as shown in Figure 1 (left). For large ${\cal T}$, we have \cite{az1} 

\begin{equation}\label{wilson}
\langle\,W({\cal C})\,\rangle \simeq \ep^{-E{\cal T}-2S}	
\,,
\end{equation}
where $E$ is a ground state energy of a quark-antiquark bound state and $S$ is a renormalized area of the end-surface shown in Figure 1 (right). 

On the other hand, in the temporal gauge the expectation value of the Wilson loop can be evaluated as \footnote{For example, see \cite{zinn} and references therein.}

\begin{equation}\label{zinn}
\langle\,W({\cal C})\,\rangle\simeq \langle\psi\vert \ep^{-H{\cal T}}\vert\psi\rangle =\ep^{-E{\cal T}}\Psi^2(r)
\,,
\end{equation}
where $H$ is the gauge Hamiltonian in the temporal gauge and $\vert\psi\rangle$ is the quark-antiquark bound state. It is given by $\vert\psi\rangle=
\Psi(r):\bar q(r)U_P(r,0)q(0):\vert 0\rangle$ such that $\langle 0\vert:\bar q (r)U_P(r,0) q(0):\vert 0\rangle =1$. Since the quarks are  
non-dynamic, we need to normal order only the $P$-exponential function that results in non-zero expectation value for the normal ordered operator. Note that the exponential fall off \eqref{f2large} is consistent with the fact that $\vert\psi\rangle$ is normalizable.

Then combining this with \eqref{wilson}, we find 

\begin{equation}\label{Wloop}
\Psi(r)\simeq\ep^{-S}
\,,
\end{equation}
where $S$ is a renormalized area of the surface shown in Figure 1 (right). Note that this formula is valid only for the ground state.

It is worth noting that a similar representation has been suggested in \cite{az3} for the expectation value of the Polyakov loop. In this case there is 
no need for the quark sources. The string is prevented from getting deep into $z$ direction by a black hole geometry such that the maximum value of $z$ is bounded by the black hole horizon. In the model with dynamic quarks it appeared in \cite{kutasov}. A difference here is that the Wilson loop goes along an internal direction. In general, it is natural to expect that in the presence of dynamic quarks the representation \eqref{Wloop} breaks down 
at large separations due to string breaking. Physically, this means that quark bound states decay.

To write a formal expression for the field strength correlator \eqref{F2}, let us think of the Wilson lines as forming a long, narrow rectangular loop 
in the $x$-$t$ plane, as shown in Figure 1 (left) but with small ${\cal T}$. The field strength operators are set at $(0,0)$ and $(r,0)$. The exponential fall off \eqref{F2large} for large $r$ is due to the Wilson loop expectation value. Subleading corrections come from quadratic fluctuations in the world-sheet path integral and from the field strength operators. They are expected to give a polynomial prefactor $\text{P}$ in front of the 
exponential. Taking the limit ${\cal T}\rightarrow 0$, we have

\begin{equation}\label{F2larger}
\text{D}_{\mu\nu ,\rho\tau}(r)\sim {\text P}_{\mu\nu,\rho\tau}\,\ep^{-2S}
\,,
\end{equation}
where $S$ is the same renormalized area as in \eqref{Wloop}. Note that this formula only provides the leading exponent in the large $r$ limit.

In analyzing the formulas \eqref{Wloop} and \eqref{F2larger}, an interesting relation arises. Since $S$ is proportional to $r$, we find that the correlation lengths $\xi$ and $\lambda$ are related as

\begin{equation}\label{cor-lengths}
\lambda=\frac{1}{2}\xi
\,.
\end{equation}

Given the background metric, we can calculate the renormalized area $S$ by using the exact shape of the static string stretched between the 
heavy quark sources \cite{az1}. With the large $r$ behavior in mind, it is technically suitable to add two pieces, shown in dashed lines in 
Figure 1 (right), to the original surface. Their areas are each of subleading order in $1/r$. As a result, the surface of interest becomes
a rectangular in the $x$-$z$ plane. 

Now we are ready to use the Nambu-Goto action equipped with the background metric \eqref{metric}

\begin{equation}\label{NG}
S=\frac{1}{2\pi\alpha'}\int d^2\tau\,\sqrt{\det \,{\cal G}_{NM}^{}\partial_\alpha X^N\partial_\beta X^M\vphantom{\bigl(\bigr)}}
\,.
\end{equation}
Next, we choose $\tau^1=x$ and $\tau^2=z$. This yields

\begin{equation}\label{NG1}
	S=\frac{\g}{\pi}r\int_0^{\zc}dz\,w=\frac{\g}{\pi}\frac{r}{\zc}\int_0^1du\,w(u)
	\,,
\end{equation}
where $\g=\tfrac{R^2}{2\alpha'}$ and $\zc=\tfrac{1}{\sqrt{\s}}$. We have also rescaled $z$ as $z=\zc u$. Since the integral is divergent at $u=0$ 
due to the $z^{-2}$ factor in the metric, we 
have to regularize it. We do so by imposing a cutoff $\epsilon$.

To next-to-leading order in $\epsilon$, the integral is given by

\begin{equation}\label{integral}
\int_\epsilon^1 du \,w=\frac{1}{\epsilon}+
\sqrt{\pi}\text{Erfi}(1)-\ep
+O(\epsilon)
\,,
\end{equation}
where $\text{Erfi}(x)$ is the imaginary error function. We use the modified minimal subtraction scheme to deal with this integral.\footnote{Note that 
the use of the modified minimal subtraction scheme in \cite{az1} allows one to adjust the value of a constant in the potential. Moreover, it also helps 
in finding a relation between the gauge/string duality result of \cite{p-loop} and the lattice data for the renormalized Polyakov loop.} So, we subtract the power divergence $1/\epsilon$ together with a constant $c$ whose value must be specified from renormalization conditions. As a result, the renormalized area takes 
the form\footnote{Of course, the constant part being a number $(\sqrt{\pi}\text{Erfi}(1)-\ep )$ can be absorbed into $c$. However, it becomes temperature-dependent at finite temperature (as we will see in section 3), so we keep it.}

\begin{equation}\label{NG3}
	S=\frac{\g}{\pi}\sqrt{\s}\,r\bigl(c+\sqrt{\pi}\text{Erfi}(1)-\ep\bigr)
\,.
\end{equation}
Combining this with \eqref{f2large} and \eqref{Wloop}, we get the correlation length $\xi$

\begin{equation}\label{lf}
	\xi=\frac{\pi}{\g\sqrt{\s}}\bigl(c+\sqrt{\pi}\text{Erfi}(1)-\ep\bigr)^{-1}
	\,.
\end{equation}

It is now clear that ${\cal T}$ must go to zero faster than $\epsilon$. Indeed, in the limit ${\cal T}/\epsilon\rightarrow 0$ the contribution of the 
lateral surface vanishes and, as a consequence, the relation \eqref{F2larger} holds. With this fact understood, it is now straightforward to find 
the correlation length $\lambda$

\begin{equation}\label{lambda}
	\lambda=\frac{\pi}{2\g\sqrt{\s}}\bigl(c+\sqrt{\pi}\text{Erfi}(1)-\ep\bigr)^{-1}
	\,.
\end{equation}

To define the model properly, we must specify the renormalization conditions. In doing so, we start with the correlation length $\xi$. Our stringy construction suggests a natural condition $\xi=1/\sqrt{\sigma}$, where $\sigma$ is the string tension. In the model we are considering 
it is given by $\sigma=\ep\g\s /\pi$ \cite{az1}. Making an estimate requires some numerics. For $SU(3)$ a value of $\g$ fixed from the heavy quark potential is $\g\approx 0.62$ \cite{white}. This gives $c\approx 3.50$. A typical value of $\s$ is $\s\approx 0.45\,\text{GeV}^2$ \cite{white,q2}. If so, a value of $\lambda$ is estimated to be 
\begin{equation}\label{estimate}
\lambda\approx 0.20\,\text{fm}
\,. 
\end{equation}
Is it a reasonable value? Although the lattice calculations of \cite{giacomo1} claim a slightly bigger value $\lambda\approx 0.22\,\text{fm}$ but with an error of order $0.03\,\text{fm}$, there is an obvious troublesome question. If the calculations are made in two different renormalization schemes, why are the results so similar? Unfortunately, we have no real resolution of this problem.  

On the other hand, choosing $\lambda=m^{-1}_{\text{\tiny G}}$, with $m_G\approx 3.64\sqrt{\sigma}$ the mass of the lightest glueball \cite{glueball}, gives $c\approx 5.62$. A simple algebra shows that in this 
case $\lambda\approx 0.13\,\text{fm}$.

\section{Calculating the Correlators at Finite Temperature}

According to \cite{review}, at finite temperature we must separately consider the electric and magnetic correlators \cite{review}. So, we decompose $G_{\mu\nu}$ into the electric and magnetic fields: $E_i=G_{0i}$ and $B_i=\tfrac{1}{2}\varepsilon_{ijk}G_{jk}$. It is expected \cite{simonov} that at large separations the magnetic correlators show exponential fall off for any temperature, while the electric ones do so just below the critical temperature $T_c$.

To study the magnetic correlator $D^{(m)}_{ij}(x)=\langle\,\text{tr}\bigl[B_i(0,\vec x)U_P(x,0)B_j(0,0)U_P(0,x)\bigr]\,\rangle$, we take the Wilson lines in the $x$-$y$ plane and regard ${\cal C}$ as a long, narrow rectangle similar to that of Figure 1 (with $t$ replaced by $y$). Thus, ${\cal C}$ is now a spatial Wilson loop. A crucial difference from temporal Wilson loops is that in the deconfined phase spatial ones obey an area law, with 
a spatial string tension $\sigma_s$.\footnote{Recall that $\sigma_s$ is not a physical string tension because it is not related to the properties of a physical potential.} In other words, a (spatial) string stretched between two well-separated sources doesn't break. This allows us to use a 
formalism similar to that of section 2. So, the exponential fall off for large $r$ is due to the spatial Wilson loop expectation value. The magnetic 
field operators and world-sheet fluctuations contribute to a polynomial prefactor $P$ in front of the exponential. Taking the ${\cal T}\rightarrow 0$ limit, we find

\begin{equation}\label{mlarger}
\text{D}_{ij}^{(m)}(r)\sim {\text P}_{ij}\,\ep^{-2S}
\,,
\end{equation}
where $S$ is a renormalized area of the surface shown in Figure 1 (right). 

Repeating the arguments of section 2, we add two pieces to the original surface to simplify further calculations. The Nambu-Gotto action, which we 
wrote before as \eqref{NG}, is now 

\begin{equation}\label{NGT}
S=\frac{1}{2\pi\alpha'}\int d^2\tau\,\sqrt{\det \,\mathfrak{G}_{NM}^{}\partial_\alpha X^N\partial_\beta X^M\vphantom{\bigl(\bigr)}}
\,,
\end{equation}
where the background metric $\mathfrak{G}$ is given by \eqref{metricT}. Using the gauge $\tau^1=x$ and $\tau^2=z$, it becomes

\begin{equation}\label{NG1T}
	S=\frac{\g}{\pi}r\int_0^{\zt}dz\,\frac{w}{\sqrt{f}}=\frac{\g}{\pi}\frac{r}{\zt}\int_0^1 dv\,\frac{w}{\sqrt{f}}(v)
	\,.
\end{equation}
Here we have rescaled $z$ as $z=\zt v$. As in section 2, we regularize the integral over $v$ by imposing a cutoff $\epsilon$. Note that in the deconfined phase the large distance physics is determined by the near horizon geometry of the deformed Schwarzschild black hole in $\text{AdS}_5$ space \cite{az2,az3}. So, the upper limit is $\zt$.

To next-to-leading order in $\epsilon$, the integral is given by 

\begin{equation}\label{NG1Ta}
	\int_\epsilon^1 dv\,\frac{w}{\sqrt{f}}=\frac{1}{\epsilon}-\sqrt{\pi}\frac{\Gamma(\frac{3}{4})}{\Gamma(\frac{1}{4})}
	{}_1F_2\bigl(-\tfrac{1}{4};\tfrac{1}{4},\tfrac{1}{2};\tfrac{1}{4}\tfrac{\T^2}{T^2}\bigr)
	+\sqrt{\pi}\frac{\T^2}{T^2}\frac{\Gamma(\frac{5}{4})}{\Gamma(\frac{3}{4})}
	{}_1F_2\bigl(\tfrac{1}{4};\tfrac{3}{4},\tfrac{3}{2};\tfrac{1}{4}\tfrac{\T^2}{T^2}\bigr)+O(\epsilon)
	\,,
	\end{equation}	
where ${}_1F_2(a;b,c;x)$ is the generalized hypergeometric function and $\T$ is given by $\T=\sqrt{\s}/\pi$ \cite{az2}. We use again the 
modified minimal subtraction scheme and look for a renormalized area. So, we have 

\begin{equation}\label{NG2T}
	S=\g\,Tr 
	\biggl[c-
	\sqrt{\pi}\frac{\Gamma(\frac{3}{4})}{\Gamma(\frac{1}{4})}
	{}_1F_2\bigl(-\tfrac{1}{4};\tfrac{1}{4},\tfrac{1}{2};\tfrac{1}{4}\tfrac{\T^2}{T^2}\bigr)
	+\sqrt{\pi}\frac{\T^2}{T^2}\frac{\Gamma(\frac{5}{4})}{\Gamma(\frac{3}{4})}
	{}_1F_2\bigl(\tfrac{1}{4};\tfrac{3}{4},\tfrac{3}{2};\tfrac{1}{4}\tfrac{\T^2}{T^2}\bigr)
	\biggr]
	\,.
\end{equation}

In the limit ${\cal T}/\epsilon\rightarrow 0$ the contribution of the lateral surface is negligible. As a result, the correlation length is 
given by 

\begin{equation}\label{lambdas}
	\lambda_s=\frac{\pi}{2\g\sqrt{\s}}\frac{\T}{T}
	\biggl[c-
	\sqrt{\pi}\frac{\Gamma(\frac{3}{4})}{\Gamma(\frac{1}{4})}
	{}_1F_2\bigl(-\tfrac{1}{4};\tfrac{1}{4},\tfrac{1}{2};\tfrac{1}{4}\tfrac{\T^2}{T^2}\bigr)
	+\sqrt{\pi}\frac{\T^2}{T^2}\frac{\Gamma(\frac{5}{4})}{\Gamma(\frac{3}{4})}
	{}_1F_2\bigl(\tfrac{1}{4};\tfrac{3}{4},\tfrac{3}{2};\tfrac{1}{4}\tfrac{\T^2}{T^2}\bigr)
	\biggr]^{-1}
	\,.
\end{equation}

To complete the picture, we must impose the renormalization conditions. Let us try $\lambda_s=1/(2\sqrt{\sigma_s})$. The reason for this is that at zero temperature it is equivalent to $\xi=1/\sqrt{\sigma}$ because the spatial string tension remains constant below $\T$, where it coincides with the 
(physical) string tension. In the model we are considering the temperature-dependence of the spatial string tension for $T>\T$ is given by \cite{az2}

\begin{equation}\label{s-tension}
\sigma_s=\sigma\frac{T^2}{\T^2}\exp\biggl\{\frac{\T^2}{T^2}-1\biggr\}	
\,,
\end{equation}
where $\sigma$ is the string tension. Note that it is in a good agreement with the lattice data up to $3-4\,\T$. 

A little experimentation with Mathematica soon shows that the expression \eqref{lambdas}, looking somewhat awkward, provides a very similar temperature dependence as the simpler expression \eqref{s-tension}. In Figure 2 we have plotted the results.  

\begin{figure}[ht]
\centering
\includegraphics[width=6.3cm]{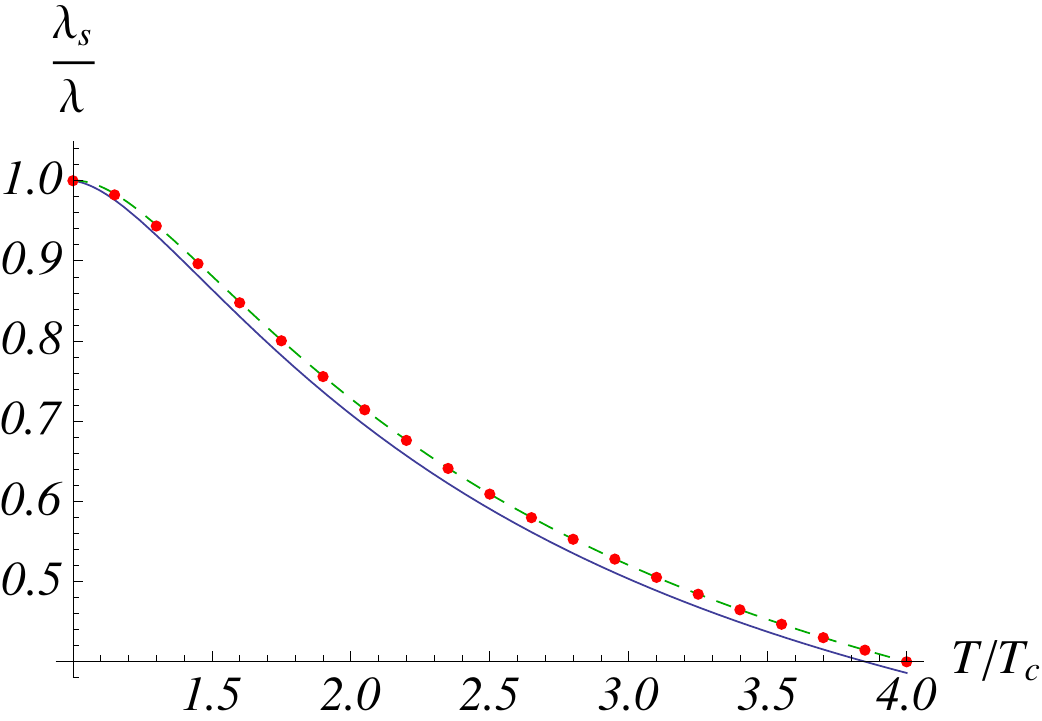}
\hfill
\includegraphics[width=6.3cm]{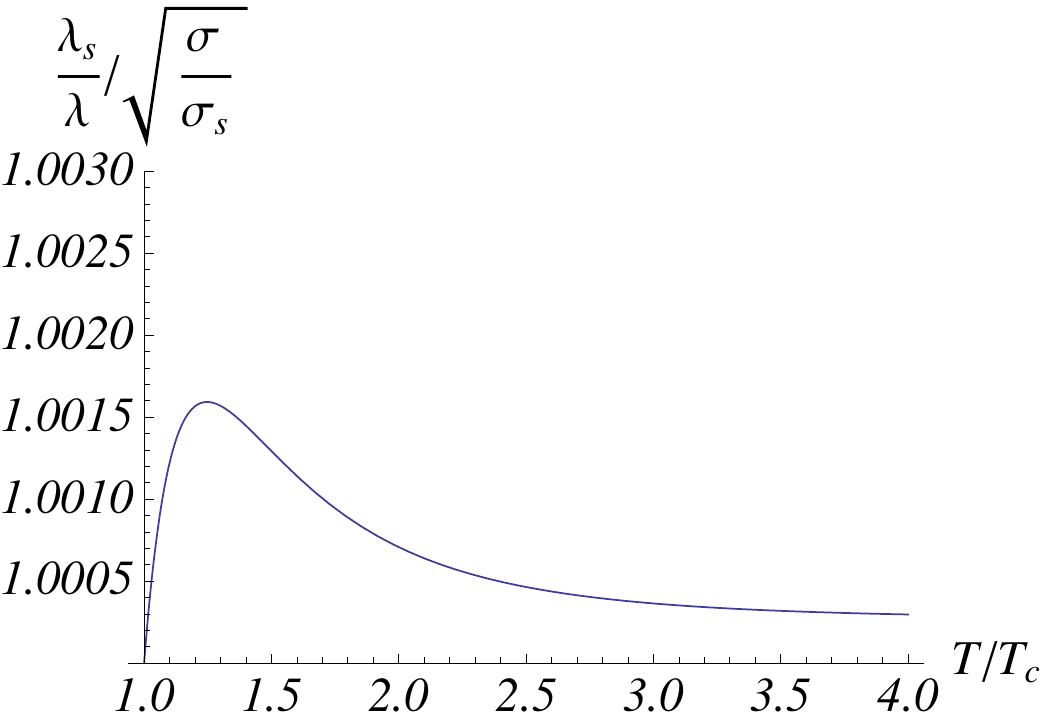}
\caption{\small{Left: A comparison of different $\frac{\lambda_s}{\lambda}$ curves for $SU(N)$ gauge theory. The solid blue curve corresponds 
to \eqref{lambdas} with $c\approx 3.50$ fixed at zero temperature. The dashed green curve represents the "best fit" with $c\approx 3.22$. In both cases, $\lambda=\lambda_s\vert_{T=\T}$. The red dots denote $\sqrt{\frac{\sigma}{\sigma_s}}$, with $\sigma_s$ given by \eqref{s-tension}.
Right: A ratio $\frac{\lambda_s}{\lambda}/\sqrt{\frac{\sigma}{\sigma_s}}$ for $c\approx 3.22$.  }} 
\end{figure}

Thus, we have 

\begin{equation}\label{sl}
\frac{\lambda_s}{\lambda}\approx\sqrt{\frac{\sigma}{\sigma_s}}
\,,
\end{equation}
with the previous value $c\approx 3.50$. It is remarkable that for $c\approx 3.22$ the maximum discrepancy is of order $0.2\%$.

To study the electric field correlator $D^{(e)}_{ij}(x)=\langle\,\text{tr}\bigl[E_i(0,\vec x)U_P(x,0)E_j(0,0)U_P(0,x)\bigr]\,\rangle$, we take the Wilson lines in the 
$x$-$t$ plane and regard them as forming a long, narrow rectangle shown in Figure 1. In the deconfined phase a static string stretched between two 
well-separated sources breaks. As a result, the dominant surface is given by a sum of two disconnected surfaces \cite{az3} whose renormalized areas vanish in the limit ${\cal T}\rightarrow 0$. This shows that in the deconfined phase there is no exponential fall off behavior of the electric field 
correlators at large distances. Such a conclusion looks attractive because it agrees with that of lattice simulations \cite{giacomo2}. 


\section{Concluding Comments}

(i) We would like to emphasize again that the field strength correlators \eqref{F2} are scheme and path dependent. The former is due to a linear divergency of expectation values of Wilson loops. It is known as the perimeter law. The latter becomes clear if we let ${\cal C}$ be a circular loop of 
radius $r/2$. This leads to $S\sim r^2$ rather than $S\sim r$ at large $r$. Note that when a value of $\cal T$ is finite, the contribution of the 
lateral surface is no longer negligible. It matters for $\lambda$ in \eqref{F2large}. 

\noindent (ii) It is a common wisdom that magnetic and electric correlators are related to spatial and temporal Wilson loops, respectively. We have followed this philosophy in section 3 with our proposal for the two-point correlators.
  
\noindent (iii) Because of scheme ambiguities, the correlation length may look not so good. It is therefore interesting that the difference 

\begin{equation}\label{mag}
\frac{\lambda_s^{-1}(T_1)}{T_1}-\frac{\lambda_s^{-1}(T_2)}{T_2}=2\g\int_0^1\frac{dv}{v^2}\bigl(1-v^4\bigr)^{-\tfrac{1}{2}}
\biggl(\exp\Bigl\{\frac{T_c^2}{T^2_1}v^2\Bigr\}-\exp\Bigl\{\frac{T_c^2}{T^2_2}v^2\Bigr\}\biggr)
\,
\end{equation}
is finite. Performing the integral gives

\begin{equation}\label{mag2}
	\begin{split}
\frac{\lambda_s^{-1}(T_1)}{T_1}-\frac{\lambda_s^{-1}(T_2)}{T_2}=
2\sqrt{\pi}\g 
\biggl[ & \frac{\Gamma(\frac{3}{4})}{\Gamma(\frac{1}{4})}
	\biggl({}_1F_2\bigl(-\tfrac{1}{4};\tfrac{1}{4},\tfrac{1}{2};\tfrac{1}{4}\tfrac{\T^2}{T_2^2}\bigr)-
	{}_1F_2\bigl(-\tfrac{1}{4};\tfrac{1}{4},\tfrac{1}{2};\tfrac{1}{4}\tfrac{\T^2}{T_1^2}\bigr)\biggr) \\
	+&\frac{\Gamma(\frac{5}{4})}{\Gamma(\frac{3}{4})}
	\biggl(\frac{\T^2}{T_1^2}{}_1F_2\bigl(\tfrac{1}{4};\tfrac{3}{4},\tfrac{3}{2};\tfrac{1}{4}\tfrac{\T^2}{T_1^2}\bigr)-
	\frac{\T^2}{T_2^2}{}_1F_2\bigl(\tfrac{1}{4};\tfrac{3}{4},\tfrac{3}{2};\tfrac{1}{4}\tfrac{\T^2}{T_2^2}\bigr)\biggr)
	\biggr]
\,.
\end{split}
\end{equation}
Thus, our model predicts the scheme-independent relation between the correlation lengths of magnetic operators at different temperatures. It will be interesting to see whether it will match or close to match lattice simulations.

\noindent (iv) If we consider not a field strength but an arbitrary local operator in the adjoint representation of $SU(N)$, then it follows from our proposal that the correlator shows the exponential fall off 

\begin{equation}\label{univ}
	\langle\,{\cal O}(0)U_P(0,x){\cal O}(x)U_P(x,0)\,\rangle \sim \ep^{-\frac{r}{\lambda}}
\qquad\text{for}\qquad r\rightarrow\infty\,,
\end{equation}
with a universal correlation length $\lambda$.

\vspace{.25cm}
{\bf Acknowledgments}

\vspace{.25cm}
This work was supported in part by DFG within the Emmy-Noether-Program under Grant No.HA 3448/3-1 and the Alexander von Humboldt Foundation under Grant No.PHYS0167. We would like to thank P.de Forcrand, M. Haack, A.I. Vainstein, V.I. Zakharov, and especially P. Weisz for useful discussions and comments. 
We also acknowledge the warm hospitality at Yukawa Institute for Theoretical Physics, where a portion of this work was done.


\end{document}